\documentclass[10pt,final,journal,twocolumn,letterpaper]{IEEEtran} 

\usepackage[dvips]{graphicx}
\usepackage{amsmath}
\usepackage{amssymb}
\usepackage[latin1]{inputenc}
\usepackage{times}
\usepackage{cite}
\usepackage{amsfonts}
\usepackage[active]{srcltx}
\usepackage{url}
\usepackage{picinpar}

\DeclareGraphicsExtensions{pst,eps}
\newcommand{\ket}[1]{| #1 \rangle}

\newcommand{\eqr}[1]{Eq.~(\ref{#1})}
\newcommand{\fref}[1]{Fig.~\ref{#1}}
\newcommand{\tref}[1]{Table~\ref{#1}}
\ifCLASSINFOpdf
\else
\fi
\input epsf
\begin{document}
%
\title{EXIT-Chart Aided Near-Capacity \\ Quantum Turbo Code Design}
\author{Zunaira Babar, Soon Xin Ng and Lajos Hanzo
\thanks{Z. Babar, S. X. Ng, and L. Hanzo are with the School of Electronics and Computer Science, University of Southampton, SO17 1BJ, United Kingdom. Email: \{zb2g10,sxn,lh\}@ecs.soton.ac.uk.}%
\thanks{The financial support of the EPSRC under the grant EP/L018659/1, that of RC-UK under the India-UK Advanced Technology Centre (IU-ATC), 
of the EU under the CONCERTO project and that of the European Research Council, Advanced Fellow Grant and of the Royal Society's Wolfson Reasearch Merit Award is gratefully acknowledged.}
}
\maketitle

\begin{abstract} 
High detection complexity is the main impediment in future Gigabit-wireless systems. However, a quantum-based detector is capable of simultaneously detecting 
hundreds of user signals by virtue of its inherent parallel nature. This in turn requires near-capacity quantum error correction codes for protecting the 
constituent qubits of the quantum detector against the undesirable environmental decoherence.
In this quest, we appropriately adapt the conventional non-binary EXtrinsic Information Transfer (EXIT) charts for quantum turbo codes by exploiting the intrinsic
quantum-to-classical isomorphism. The EXIT chart analysis not only allows us to dispense with the time-consuming Monte-Carlo simulations, but also facilitates
the design of near-capacity codes without resorting to the analysis of their distance spectra. We have demonstrated that our EXIT chart predictions 
are in line with the Monte-Carlo simulations results. We have also optimized
the entanglement-assisted QTC using EXIT charts, which outperforms the existing distance spectra based QTCs. More explicitly, the performance of our optimized
QTC is as close as $0.3$ dB to the corresponding hashing bound. 
\end{abstract}

\begin{keywords}
Quantum Error Correction, Turbo Codes, EXIT Charts, Near-Capacity Design.
\end{keywords}

%
\IEEEpeerreviewmaketitle
\section{Introduction} \label{sec:intro}
Multi-User Multiple-Input Multiple-Output (MU-MIMO)~\cite{mimo_sxn_2006, mimo_2012} and massive MIMO~\cite{massive_mimo} schemes are promising candidates 
for the future generation 
Gigabit-wireless system. However, the corresponding detection complexity increases exponentially with the number of users and antennas, when aiming
for approaching the optimum Maximum-Likelihood (ML) performance. An attractive solution to this exponentially escalating complexity problem is to perform the 
ML detection in the quantum domain, since quantum 
computing allows parallel evaluations of a function at a complexity cost that is equivalent to a single classical evaluation~\cite{Qbook3,panos2013}. 
However, a quantum detector requires powerful Quantum Error Correction codes (QECC's) for stabilizing and protecting the fragile constituent quantum bits (qubits) 
against the undesirable quantum decoherence, when they interact with the environment~\cite{Qbook3,Qbook2}. Furthermore, quantum-based wireless transmission
is capable of supporting secure data dissemination~\cite{Qbook3,secure_2010}, where any `measurement' or `observation' by an eavesdropper will destroy the quantum 
entanglement, hence intimating the parties concerned~\cite{Qbook3}. However, this requires powerful QECC's for the reliable transmission 
of qubits across the wireless communication channels. Hence, near-capacity QECC's are the vital enabling technique for future generations of 
wireless systems, which are both reliable and secure, yet operate at an affordable detection complexity.
%

Classical turbo codes operate almost arbitrarily close to the Shannon limit, which inspired researchers to achieve a comparable near-capacity performance
for quantum systems~\cite{qturbo1,qturbo2,wilde_turbo,wilde_turbo2,zbabar2013_2}. In this quest,
Poulin \textit{et al.} developed the theory of Quantum Turbo Codes (QTCs) in~\cite{qturbo1,qturbo2}, based
on the interleaved serial concatenation of Quantum Convolutional Codes (QCCs)~\cite{ollivier2003,ollivier2004,forney2007,grassal2007},
and investigated their behaviour on a quantum depolarizing channel\footnote{A quantum channel can be used for modeling imperfections in quantum hardwares, namely faults resulting from quantum 
decoherence and quantum gates. Furthermore, a quantum channel can also model quantum-state flips imposed by the transmission medium, including free-space 
wireless channels and optical fiber links, when qubits are transmitted across these media.}.
It was found in~\cite{qturbo1,Monireh2012} that the constituent QCCs cannot be simultaneously recursive and non-catastrophic. Since recursive nature of the inner code
is essential for ensuring an unbounded minimum distance, while the non-catastrophic nature is required for achieving decoding convergence, the QTCs of~\cite{qturbo1,qturbo2}
had a bounded minimum distance. 
More explicitly, the design of Poulin \textit{et al.}~\cite{qturbo1,qturbo2} was based on non-recursive and 
non-catastrophic convolutional codes. Later, Wilde and Hseih~\cite{wilde_turbo} extended the concept of pre-shared entanglement to QTCs, which 
facilitated the design of QTCs having an unbounded minimum distance. Wilde \textit{et al.} also introduced the notion of \textit{extrinsic} information to 
the iterative decoding of QTCs
and investigated various code structures in~\cite{wilde_turbo2}.

The search for the optimal components of a QTC has been so far confined to the analysis of the constituent QCC distance spectra, followed 
by intensive Monte-Carlo simulations for determining the convergence threshold of the resultant QTC, as detailed in~\cite{qturbo2,wilde_turbo2}. 
While the distance spectrum dominates a 
turbo code's performance in the Bit Error Rate (BER) floor region, it has a relatively insignificant impact on the convergence properties in the turbo-cliff 
region~\cite{brink:exit-parallel}. Therefore, having a good distance spectrum does not guarantee having a near-capacity performance - in fact, often there is a
trade-off between them. 
To circumvent this problem and to dispense with time-consuming Monte-Carlo simulations, 
in this contribution we extend the application of EXIT charts to the design of quantum turbo codes.

\textit{More explicitly,
\begin{itemize}
 \item We have appropriately adapted the conventional non-binary EXIT chart based design approach to the family of quantum turbo codes based on the underlying 
quantum-to-classical isomorphism. Similar to the classical codes, our EXIT chart predictions are in line with the Monte-Carlo simulation results. 
\item We have analyzed the behaviour of both an un-assisted (non-recursive) and of an entanglement-assisted (recursive) inner convolutional code using EXIT 
charts for demonstrating that, similar to their classical counterparts, recursive inner quantum codes constitute families of QTCs having an unbounded minimum distance.
\item For the sake of approaching the achievable capacity, we have optimized the constituent inner and outer components of QTC using EXIT charts. In contrast 
to the distance spectra based QTCs of~\cite{wilde_turbo2} whose performance is within $0.9$ dB of the hashing bound,
our optimized QTC operates within $0.3$ dB of the capacity limit.
However, our intention was not to carry out an exhaustive code search over the potentially excessive parameter-space, but instead to demonstrate how
our EXIT-chart based approach may be involved for quantum codes.
This new design-approach is expected to stimulate further interest in the EXIT-chart based near-capacity design of various concatenated quantum codes. 
\end{itemize}}

This paper is organized as follows. Section~\ref{sec1} provides a rudimentary introduction to quantum stabilizer codes and QTCs.
We will then present our proposed EXIT-chart based approach conceived for quantum turbo codes in Section~\ref{sec2}. Our
results will be discussed in Section~\ref{sec3}, while our conclusions are offered in Section~\ref{sec4}.
\section{Preliminaries} \label{sec1}
The constituent convolutional codes of a QTC belong to the class of stabilizer codes~\cite{got2}, which are analogous to classical linear block codes.
We will here briefly review the basics of stabilizer codes in order to highlight this relationship for the benefit of readers with background in
classical channel coding. This will be followed by a brief discussion of QTCs.
\subsection{Stabilizer Codes} \label{sec1a}
Qubits collapse to classical bits upon measurement~\cite{Qbook2,panos2013}. This prevents us from directly applying classical error correction
techniques for reliable quantum transmission. Quantum error correction codes circumvent this problem by observing the error syndromes without reading the actual
quantum information. Hence, quantum stabilizer codes invoke the syndrome decoding approach of classical linear block codes
for estimating the errors incurred during transmission.

Let us first recall some basic definitions~\cite{Qbook2}.
\paragraph*{Pauli Operators} The $\mathbf{I}$, $\mathbf{X}$, $\mathbf{Y}$ and $\mathbf{Z}$ Pauli operators are defined by the following matrices:
\begin{align}
 \mathbf{I}&=\begin{pmatrix}
  1& 0 \\
  0& 1
\end{pmatrix}, \ 
\mathbf{X}=\begin{pmatrix}
  0& 1 \\
  1& 0
\end{pmatrix}, \nonumber \\ 
\mathbf{Y}&=\begin{pmatrix} 
  0& -i \\
  i& 0
\end{pmatrix}, \ 
\mathbf{Z}=\begin{pmatrix}
  1& 0 \\
  0& -1
\end{pmatrix},\label{eq:IXYZ}
\end{align}
where the $\mathbf{X}$, $\mathbf{Y}$ and $\mathbf{Z}$ operators anti-commute with each other.
\paragraph*{Pauli Group} A single qubit Pauli group $\mathcal{G}_1$ consists of all the Pauli matrices of \eqr{eq:IXYZ} together with the multiplicative factors 
$\pm 1$ and $\pm i$, i.e. we have:
\begin{equation}
 \mathcal{G}_1 \equiv \{\pm \mathbf{I}, \pm i \mathbf{I},\pm \mathbf{X}, \pm i \mathbf{X}, \pm \mathbf{Y}, \pm i \mathbf{Y}, \pm \mathbf{Z}, \pm i \mathbf{Z}\}.
\label{eq:gn}
\end{equation}
The general Pauli group $\mathcal{G}_n$ is an $n$-fold tensor product of $\mathcal{G}_1$.
\paragraph*{Depolarizing Channel}
The depolarizing channel characterized by the probability $p$ inflicts an $n$-tuple error $\mathcal{P} \in \mathcal{G}_n$ on $n$ qubits, where 
the $i^{th}$ qubit may experience either a bit flip ($\mathbf{X}$), a phase flip ($\mathbf{Z}$) or both ($\mathbf{Y}$) with a probability of $p/3$.

An $[n,k]$ Quantum Stabilizer Code (QSC), constructed over a code space $\mathcal{C}$, is defined by a set of $(n - k)$ independent commuting $n$-tuple Pauli 
operators $g_i$, for $1 \leq i \leq (n - k)$. The corresponding encoder then maps the information word (logical qubits)
$\ket{\psi} \in \mathbb{C}^{2^k}$ onto the codeword (physical qubits) $\ket{\overline{\psi}} \in \mathbb{C}^{2^n}$,
where $\mathbb{C}^d$ denotes the $d$-dimensional Hilbert space. 
More specifically, the corresponding stabilizer group $\mathcal{H}$ contains both
$g_i$ and all the products of $g_i$ for $1 \leq i \leq (n - k)$ and forms an abelian subgroup of $\mathcal{G}_n$. A unique feature of these operators is that 
they do not change the state of valid codewords, while yielding an eigenvalue of $-1$ for corrupted states. Consequently, the eigenvalue
is $-1$ if the $n$-tuple Pauli error $\mathcal{P}$ anti-commutes with the stabilizer $g_i$ and it is $+1$ if $\mathcal{P}$
commutes with $g_i$. More explicitly, we have:
\begin{equation}
 g_i \ket{\hat{\psi}} = \left\{
\begin{array}{l l}
 \ket{\overline{\psi}}, & g_i \mathcal{P} = \mathcal{P} g_i \\
-\ket{\overline{\psi}}, & g_i \mathcal{P} = - \mathcal{P} g_i, \\
\end{array}
\right.
\label{eq:stab}
\end{equation}
where $\mathcal{P}$ is an $n$-tuple Pauli error, $\ket{\overline{\psi}} \in \mathcal{C}$ and $\ket{\hat{\psi}} = \mathcal{P} \ket{\overline{\psi}}$ is the received codeword.
The resultant $\pm 1$ eigenvalue gives the corresponding error syndrome, which is $0$ for an eigenvalue of $+1$ and $1$ for an eigenvalue of $-1$. 
It must be mentioned here that Pauli errors which differ only by the stabilizer group have the same impact on all the codewords and therefore can be
corrected by the same recovery operations. This gives quantum codes the intrinsic property of degeneracy~\cite{deg_poulin}. 

As detailed in~\cite{cleve97,sparse1}, QSCs may be characterized in terms of an equivalent binary parity check matrix notation satisfying the commutativity
constraint of stabilizers. This can be exploited for designing quantum codes
with the aid of known classical codes. The $(n - k)$ stabilizers of an $[n,k]$ stabilizer code can be represented as a concatenation of a pair of 
$(n - k) \times n$ binary matrices $\mathbf{H}_z$ and $\mathbf{H}_x$, resulting in the binary parity check matrix $\mathbf{H}$ as given below:
\begin{equation}
 \mathbf{H} = [\mathbf{H}_z | \mathbf{H}_x]. 
\label{eq:eq-classical}
\end{equation}
More explicitly, each row of $\mathbf{H}$ corresponds to a stabilizer of $\mathcal{H}$, so that the $i^{th}$ column
of $\mathbf{H}_z$ and $\mathbf{H}_x$
corresponds to the $i^{th}$ qubit and a binary $1$ at these locations represents a $\mathbf{Z}$ and $\mathbf{X}$ Pauli operator, respectively, in the corresponding stabilizer. Moreover,
the commutativity requirement of stabilizers is transformed into the orthogonality of rows with respect to the symplectic product defined in~\cite{sparse1}, as follows:
\begin{equation}
 \mathbf{H}_z \mathbf{H}_{x}^{T} + \mathbf{H}_x \mathbf{H}_{z}^{T} = 0.
\label{eq:symp}
\end{equation}
Conversely, two classical linear codes $\mathbf{H}_z$ and $\mathbf{H}_x$ can be used to construct a quantum stabilizer code $\mathbf{H}$ of \eqr{eq:eq-classical} if
$\mathbf{H}_z$ and $\mathbf{H}_x$ meet the symplectic criterion of \eqr{eq:symp}.

In line with this discussion, a Pauli error operator $\mathcal{P}$ can be represented by the effective error $P$, which is a binary vector of length $2n$. 
More specifically, $P$ is a concatenation of $n$ bits for $\mathbf{Z}$ errors, followed by another $n$ bits for $\mathbf{X}$ errors and the resultant 
syndrome is given by the symplectic product of $\mathbf{H}$ and $P$, which is equivalent to $\mathbf{H}[P_x:P_z]^T$.
In other words, the Pauli-$\mathbf{X}$ operator is used for correcting $\mathbf{Z}$ errors, while the Pauli-$\mathbf{Z}$ operator is used for correcting $\mathbf{X}$ errors~\cite{Qbook2}.
Thus, the quantum-domain syndrome is equivalent to the classical-domain binary syndrome 
and a basic quantum-domain decoding procedure is similar to syndrome based decoding of the equivalent classical code~\cite{sparse1}. 
However, due to the degenerate nature of quantum codes, quantum decoding aims for finding the most likely error coset, while the classical syndrome decoding finds the most likely error. 

\subsection{Quantum Turbo Codes} \label{sec1b}
Analogous to classical Serially Concatenated (SC) turbo codes, QTCs are obtained from the interleaved serial concatenation of QCCs, which belong to the class
of stabilizer codes. However,
it is more convenient to exploit the circuit-based representation of the constituent codes, rather than the conventional parity-check matrix based syndrome
decoding~\cite{zbabar2013}.
Before proceeding with the decoding algorithm, we will briefly review the circuit-based representation. This discussion is based on~\cite{qturbo2}.

Let us consider an $(n,k)$ classical linear block code constructed over the code space $C$, which maps the information word $c \in \mathbb{F}^k_2$ onto the corresponding 
codeword 
$\overline{c} \in \mathbb{F}^n_2$. In the circuit-based representation, the code space $C$ is defined as follows:
\begin{equation}
 C = \{\overline{c} = (c:0_{n-k})V\},
\end{equation}
where $V$ is an $(n \times n)$-element invertible encoding matrix over $\mathbb{F}_2$. Similarly, for an $[n,k]$ quantum stabilizer code, the
quantum code space $\mathcal{C}$ is defined as:
\begin{equation}
 \mathcal{C} = \{\ket{\overline{\psi}} = \mathcal{V}(\ket{\psi} \otimes \ket{0_{n-k}}) \},
\end{equation}
where $\mathcal{V}$ is an $n$-qubit Clifford transformation\footnote{Clifford transformation $\mathcal{V}$ is a unitary transformation, which maps an $n$-qubit Pauli group $\mathcal{G}_n$ onto itself under conjugation~\cite{clif2003},
i.e.
\begin{equation}
 \mathcal{V} \mathcal{G}_n \mathcal{V}^\dag = \mathcal{G}_n. \nonumber
\label{eq:bin_V}
\end{equation}
}
and $\ket{\psi} \in \mathbb{C}^{2^k}$. The corresponding binary encoding matrix $V$ is a unique $(2n \times 2n)$-element matrix
such that for any $\mathcal{P} \in \mathcal{G}_n$ we have~\cite{qturbo2}: 
\begin{equation}
 [\mathcal{V} \mathcal{P} \mathcal{V}^\dag] = [\mathcal{P}]V,
\end{equation} 
where $[\mathcal{P}] = P$ and $[.]$ denotes the effective Pauli group $G_n$ such that $[\mathcal{P}]$ differs from $\mathcal{P}$ by a multiplicative constant, i.e. we have $[\mathcal{P}] = \mathcal{P}/\{\pm1,\pm\textit{i}\}.$
The rows of $V$, denoted as $V_i$ for $1\leq i \leq 2n$, are given by $V_{i} = [\mathcal{V} Z_i \mathcal{V}^\dag] = [Z_i]V$ for $1\leq i \leq n$ 
and $V_{i} = [\mathcal{V} X_i \mathcal{V}^\dag] = [X_i]V$ for $n < i \leq 2n$. Here $X_i$ and $Z_i$ represents the Pauli $\mathbf{X}$ and $\mathbf{Z}$ operators acting on the $i^{th}$ qubit.
Furthermore, any codeword in $\mathcal{C}$ is invariant by $\mathcal{V} Z_i \mathcal{V}^\dag$, for $k < i \leq n$, which therefore corresponds to the
stabilizer generators $g_i$ of \eqr{eq:stab}. More explicitly, the rows $V_{i}$, for $k < i \leq n$, constitute the $(n-k) \times 2n$ parity check matrix $\mathbf{H}$
of \eqr{eq:eq-classical}, which meets the symplectic criterion of \eqr{eq:symp}. 

\begin{figure}[tb]
  \begin{center}
  {\includegraphics[width=1\linewidth]{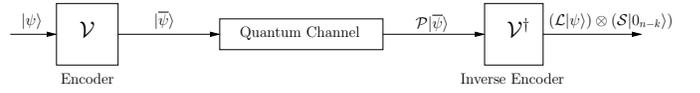}}
    \caption{Quantum transmission model.}
  \label{fig:QSC}
  \end{center}
\end{figure}

At the decoder, the received codeword $\ket{\hat{\psi}} = \mathcal{P} \ket{\overline{\psi}}$ is passed through the inverse encoder $\mathcal{V}^{\dag}$,
which yields the corrupted transmitted information word $\mathcal{L}|\psi\rangle$ and the associated syndrome $\mathcal{S}|0_{n-k}\rangle$, formulated as:
\begin{align}
 \mathcal{V}^{\dag}\mathcal{P}|\overline{\psi}\rangle &= \mathcal{V}^{\dag}\mathcal{P}\mathcal{V}(|\psi\rangle \otimes |0_{n-k}\rangle) \nonumber \\
& = (\mathcal{L}|\psi\rangle) \otimes (\mathcal{S}|0_{n-k}\rangle),
\end{align}
where $\mathcal{L}$ denotes the error imposed on the logical qubits, while $\mathcal{S}$ represents the error inflicted on the remaining $(n-k)$ qubits. 
This transmission process is summarized in \fref{fig:QSC}.

Since stabilizer codes are analogous to linear block codes, syndrome decoding is employed at the receiver to find the most likely error coset $\mathcal{L}$
given the syndrome $\mathcal{S}$. This is efficiently achieved by exploiting the equivalent binary
encoding matrix $V$ of \eqr{eq:bin_V}, which decomposes the effective $n$-qubit error imposed on the physical qubits $P = [\mathcal{P}]$ into the effective 
$k$-qubit error inflicted on the logical qubits $L = [\mathcal{L}]$ and the corresponding effective
$(n - k)$-qubit syndrome $S = [\mathcal{S}]$, as portrayed in \fref{fig:cct_V} and mathematically represented below: 
\begin{equation}
 PV^{-1} = (L : S).
\label{eq:cct}
\end{equation} 
\begin{figure}[tb]
  \begin{center}
  {\includegraphics[width=0.4\linewidth]{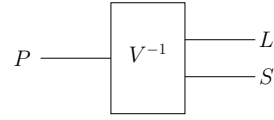}}
    \caption{Circuit representation of the inverse encoder $PV^{-1} = (L : S)$.}
  \label{fig:cct_V}
  \end{center}
\end{figure}
More explicitly, $P \in G_n$, $L \in G_k$ and $S \in G_{n-k}$.

For an $[n,k,m]$ QCC, the encoding matrix $V$ is constructed from repeated uses of the seed transformation $U$ 
shifted by $n$ qubits, as shown in Fig. $6$ of~\cite{qturbo2}. More specifically, $U$ is the binary equivalent of an $(n+m)$-qubit symplectic matrix. 
Furthermore, \eqr{eq:cct} may be modified as follows~\cite{qturbo2}:
\begin{equation}
 (P_t:M_t)U^{-1} = (M_{t-1}:L_t:S_t),
\label{eq:cct_U}
\end{equation}
where $t$ and $(t-1)$ denotes the current and previous time instants, respectively, while $M$ is the effective $m$-qubit error on the memory states.
Furthermore, $2(n-k)$-element binary vector $S$ of \eqr{eq:cct} and \eqref{eq:cct_U} can be decomposed into two components, yielding $S = S^x + S^z$, wher $S^x$ and $S^z$ are the $\mathbf{X}$ and $\mathbf{Z}$
components of the syndrome $S$, respectively. The $(n-k)$-binary error syndrome computed using the parity check matrix $\mathbf{H}$ only
reveals $S^x$ but not $S^z$~\cite{qturbo2}. Therefore, those physical errors which only differ in $S^z$ do not have to be differentiated,
since they correspond to the same logical error $L$ and
can be corrected by the same operations. These are the degenerate errors, which only differ by the stabilizer group as discussed in Section~\ref{sec1a}.
Consequently, a quantum turbo decoding algorithm aims for finding the most likely error coset acting on the logical qubits,
i.e. $L$, which satisfies the syndrome $S^x$. 

\begin{figure}[tb]
  \begin{center}
  {\includegraphics[width=1\linewidth]{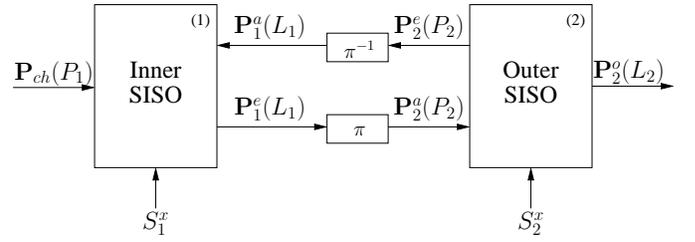}}
    \caption{Schematic of the quantum turbo decoder. \textit{$\mathbf{P}^a_i(.)$, $\mathbf{P}^e_i(.)$ and $\mathbf{P}^o_i(.)$ are the \textit{a-priori}, \textit{extrinsic}
and \textit{a-posteriori}
probabilities related to the $i^{th}$ decoder; $P_i$ and $L_i$ denote the error on the physical and logical qubits, while $S^x_i$ represents the syndrome sequence
for the $i^{th}$ decoder.}}
  \label{fig:turbo_dec}
  \end{center}
\end{figure}

Similar to the classical turbo codes, Quantum turbo decoding invokes an iterative decoding algorithm at the receiver for exchanging \textit{extrinsic} information~\cite{wilde_turbo2,tc_teq_st_2:book}
between the pair of SC Soft-In Soft-Out (SISO) decoders, as shown in \fref{fig:turbo_dec}.
These SISO decoders employ the degenerate decoding approach of~\cite{qturbo2}. Let $P_i$ and $L_i$ denote the error imposed on the physical and logical qubits, 
while $S^x_i$ represents the syndrome sequence for the $i^{th}$ decoder. Furthermore, $\mathbf{P}^a_i(.)$, $\mathbf{P}^e_i(.)$ and $\mathbf{P}^o_i(.)$ denote
the \textit{a-priori}, \textit{extrinsic} and \textit{a-posteriori} probabilities~\cite{tc_teq_st_2:book} related to the $i^{th}$ decoder. Based on this notation, the turbo
decoding process can be summarized as follows: 
\begin{itemize}
\item The inner SISO decoder uses the channel information $\mathbf{P}_{ch}(P_1)$, the \textit{a-priori} information gleaned from the outer decoder $\mathbf{P}_1^{a}(L_1)$ (initialized
to be equiprobable for the first iteration) and the syndrome $S^x_{1}$ to compute the \textit{extrinsic} information $\mathbf{P}_1^{e}(L_1)$.
\item $\mathbf{P}_1^{e}(L_1)$ is passed through a quantum interleaver\footnote{An $N$-qubit quantum interleaver is an $N$-qubit 
symplectic transformation, which randomly permutes the $N$ qubits and also applies single-qubit symplectic transformations to the individual 
qubits~\cite{qturbo2}.} $(\pi)$ to yield \textit{a-priori} information for the outer decoder $\mathbf{P}_2^{a}(P_2)$.
\item Based on the \textit{a-priori} information $\mathbf{P}_2^{a}(P_2)$ and on the syndrome $S^x_{2}$, the outer SISO decoder computes both
the \textit{a-posteriori} information $\mathbf{P}_2^{o}(L_2)$ and the \textit{extrinsic} information $\mathbf{P}_2^{e}(P_2)$.
\item $\mathbf{P}_2^{e}(P_2)$ is de-interleaved to obtain $\mathbf{P}_1^{a}(L_1)$, which is fed back to the inner SISO decoder. This iterative procedure continues
until convergence is achieved or the maximum affordable number of iterations is reached.
\item Finally, a qubit-based MAP decision is made to determine the most likely error coset $L_2$.
\end{itemize}
\section{Application of EXIT Charts to Quantum Turbo Codes} \label{sec2} 
In this section, we will extend the application of EXIT charts to the quantum domain, by appropriately
adapting the conventional non-binary EXIT chart generation technique to the circuit-based quantum syndrome decoding approach. 
Some of the information presented in this section might seem redundant to the experts of classical channel coding theory. 
However, since EXIT charts are not widely known in the quantum community, this introduction was necessary to make this treatise
accessible to quantum researchers.
\subsection{EXIT Charts} \label{sec2:exit}
EXIT charts~\cite{brink:exit-parallel,tc_teq_st_2:book,Hajjar2013} are capable of visualizing the convergence behaviour of iterative decoding
schemes by exploiting the input/output relations of the constituent decoders in terms of their average Mutual Information (MI) characteristics. They have been extensively employed for designing near-capacity
classical codes~\cite{near_cap_brink,near_cap_2010,near_cap_2007}. 
Let us recall that the EXIT chart of a serially concatenated scheme visualizes the exchange of 
the following four MI terms:
\begin{itemize}
 \item average \textit{a-priori} MI of the inner decoder, $I_A^1$,
 \item average \textit{a-priori} MI of the outer decoder, $I_A^2$,
 \item average \textit{extrinsic} MI of the inner decoder, $I_E^1$, and
 \item average \textit{extrinsic} MI of the outer decoder, $I_E^2$.
\end{itemize}
More specifically, $I_A^1$ and $I_E^1$ constitute the EXIT curve of the inner decoder, while $I_A^2$ and $I_E^2$ yield the EXIT curve of the outer
decoder. The MI transfer characteristics of both the decoders are plotted in the same graph, with the $x$ and $y$ axes
of the outer decoder swapped. The resultant EXIT chart quantifies the improvement in the mutual information as the iterations proceed,
which can be viewed as a stair-case-shaped decoding trajectory. 
Having an open tunnel between the two EXIT curves ensures that the decoding trajectory
reaches the $(1,y)$ point of perfect convergence. 
\subsection{Quantum-to-Classical Isomorphism} \label{sec2:classical_eq}
\begin{figure}[tb]
  \begin{center}
  {\includegraphics[width=0.75\linewidth]{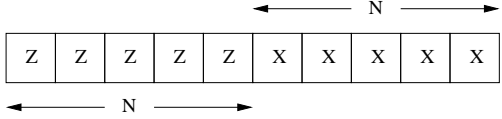}}
    \caption{Effective error $(P)$ corresponding to the error imposed on an $N$-qubit frame $(\mathcal{P})$.}
  \label{fig:eq_q_ch}
  \end{center}
\end{figure}

Before proceeding with the application of EXIT charts for quantum codes, let us elaborate on the quantum-to-classical isomorphism encapsulated in
\eqr{eq:eq-classical}, which forms the basis of our EXIT chart aided approach. 
As discussed in Section~\ref{sec1a}, a Pauli error operator $\mathcal{P}$ experienced by an $N$-qubit frame transmitted
over a depolarizing channel can be modeled by an effective error-vector $P$, which is a binary vector of length $2N$. The first $N$ bits of $P$ denote $Z$ errors,
while the remaining $N$ bits represent $\mathbf{X}$ errors, as depicted in \fref{fig:eq_q_ch}. More explicitly, an $\mathbf{X}$ error imposed on the $1^{st}$ qubit
will yield a $0$ and a $1$ at the $1^{st}$ and $(N+1)^{th}$ index of $P$, respectively. Similarly, a $\mathbf{Z}$ error imposed on the $1^{st}$ qubit
will give a $1$ and a $0$ at the $1^{st}$ and $(N+1)^{th}$ index of $P$, respectively, while a $Y$ error on the $1^{st}$ qubit will
result in a $1$ at both the $1^{st}$ as well as $(N+1)^{th}$ index of $P$. Since a depolarizing channel characterized by the probability $p$ incurs
$\mathbf{X}$, $\mathbf{Y}$ and $\mathbf{Z}$ errors with an equal probability of $p/3$, the effective error-vector $P$ reduces to two Binary Symmetric Channels 
(BSCs) having a crossover probability of $2p/3$, where we have one channel for the $\mathbf{Z}$ errors and the other for the $\mathbf{X}$ errors.
Hence, a quantum depolarizing channel has been considered analogous to a Binary Symmetric Channel (BSC)~\cite{sparse1,mark:book}, whose capacity is given by:
\begin{equation}
 C_{\text{BSC}} = 1 - H_2(2p/3),
\end{equation}
where $H_2$ is the binary entropy function. Using \eqr{eq:eq-classical}, we can readily infer that the code rate $R_Q$ of an $[n,k]$ QSC is related to the equivalent classical code rate 
$R_C$ as follows~\cite{sparse1,tan2010}:
\begin{equation}
 R_C = \frac{1}{2} \left(1 + R_Q\right).
\label{eq:rc}
\end{equation}
Consequently, the corresponding quantum capacity is as follows~\cite{sparse1,tan2010}:
\begin{equation}
C^{Q}_{\text{BSC}} = 1 - H_2(2p/3).
\end{equation}
\begin{table}[t]
 \centering
\begin{center}
\begin{tabular}{c|c|c}
\hline
 & $\mathbf{Z} = 0$ & $\mathbf{Z} = 1$\\ 
\hline
$\mathbf{X} = 0$ & $1 - p$ & $p/3$\\ 
$\mathbf{X} = 1$ & $p/3$ & $p/3$\\
\hline
 \end{tabular}
 \end{center}
 \caption{Correlation between $\mathbf{X}$ and $\mathbf{Z}$ errors on th $i^{th}$ qubit in terms of the corresponding probability of occurrence.}
 \label{tab:cor_p}
\end{table}
However, the two BSCs constituting a quantum depolarizing channel are not entirely independent. There is an inherent correlation
between the $\mathbf{X}$ and $\mathbf{Z}$ errors~\cite{sparse1}, which is characterized in \tref{tab:cor_p}. This correlation is taken into account by the 
turbo decoder of \fref{fig:turbo_dec}. 
Alternatively, a quantum depolarization channel can also be considered equivalent to a $4$-ary symmetric channel. More explicitly, the $i^{th}$
and $(N+i)^{th}$ index of $P$ constitute the $4$-ary symbol. The corresponding classical capacity is equivalent to the maximum rate achievable
over each half of the $4$-ary symmetric channel, as follows~\cite{sparse1,tan2010}:
\begin{equation}
 C_{\text{4-ary}} = \frac{1}{2}[2 - H_2(p) - p\log_2(3)].
\label{eq:c4}
\end{equation}
Therefore, using \eqr{eq:rc} the corresponding quantum capacity can be readily shown to be~\cite{sparse1,tan2010}:
\begin{equation}
 C^{Q}_{\text{4-ary}} = 1 - H_2(p) - p\log_2(3),
\end{equation}
which is known as the hashing bound\footnote{Hashing bound determines the code rate at which a random quantum code facilitates reliable
transmission for a particular depolarizing probability $p$~\cite{wilde_turbo2}.}.

Recall that a quantum code is equivalent to a classical code through \eqr{eq:eq-classical}. More specifically, as mentioned in Section~\ref{sec1a}, the decoding of a quantum code is essentially 
carried out with the aid of the equivalent classical code by exploiting the additional property
of degeneracy. Quantum codes employ syndrome decoding~\cite{zbabar2013}, which yields information about the \textbf{error-sequence} rather than the information-sequence 
or coded qubits, hence avoiding the observation of the latter sequences, which would collapse them back to the classical domain.

Since a depolarizing channel is analogous to the BSC and a QTC has an equivalent classical representation, we employ the EXIT chart technique to design near-capacity QTCs. 
The major difference between the EXIT charts conceived for the classical and quantum domains is that while the former models the \textit{a-priori}
information concerning the input bits of the inner encoder (and similarly the output bits of the outer encoder), the latter models the
\textit{a-priori} information concerning the corresponding error-sequence, i.e. the error-sequence related to the input qubits of the inner encoder $L_1$
(and similarly error-sequence related to the output qubits of the outer encoder $P_2$). This will be dealt with further in the next section.
\subsection{EXIT Charts for Quantum Turbo Codes} \label{sec2:system}
Similar to the classical EXIT charts, in our design we assume that
the interleaver length is sufficiently high to ensure that~\cite{brink:exit-parallel,tc_teq_st_2:book}:
\begin{itemize}
 \item the \textit{a-priori} values are fairly uncorrelated; and
 \item the \textit{a-priori} information has a Gaussian distribution.
\end{itemize}
 \begin{figure}[tb]
  \begin{center}
  {\includegraphics[width=1\linewidth]{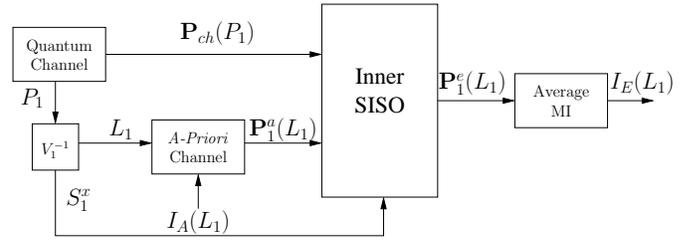}}
    \caption{System model for generating the EXIT chart of the inner decoder.}
  \label{fig:exit_inner}
  \end{center}
\end{figure}
\fref{fig:exit_inner} shows the system model used for generating the EXIT chart of the inner decoder. Here, a quantum
depolarizing channel having a depolarizing probability of $p$ generates the error sequence $P_1$, which is passed through the inverse inner encoder $V_1^{-1}$. This
yields both the error imposed on the logical qubits $L_1$ and the syndrome $S^x_1$ according to \eqr{eq:cct}. 
The \textit{a-priori} channel block then models the 
\textit{a-priori} information $\mathbf{P}^{a}_1(L_1)$ such that the average mutual information between the actual error $L_1$ and the \textit{a-priori} 
probabilities $\mathbf{P}^{a}_1(L_1)$ is given by $I_A(L_1)$~\cite{brink:exit-parallel,tc_teq_st_2:book,Hajjar2013}. 
More explicitly, we have $I_A(L_1) = I[L_1,\mathbf{P}^{a}_1(L_1)]$, where $I$ denotes the average mutual information function.
As discussed in Section~\ref{sec2:classical_eq}, the $i^{th}$ and $(N+i)^{th}$ bits of the effective error vector $L_1$ can be visualized as $4$-ary symbols. 
Consequently, similar to classical non-binary EXIT charts~\cite{grant2001,kliewer:exit-symbol}, the \textit{a-priori} information is modeled using an independent Gaussian distribution 
with a mean of zero and variance of $\sigma_A^2$, assuming that the $\mathbf{X}$ and $\mathbf{Z}$ errors constituting the $4$-ary symbols are independent. 
Using the channel information $\mathbf{P}_{ch}(P_1)$, syndrome $S^x_1$ and the \textit{a-priori} information, the inner SISO decoder yields the \textit{extrinsic} 
information $\mathbf{P}^{e}_1(L_1)$ based on the classic forward-backward recursive coefficients $\alpha_t$ and $\beta_t$ as follows\cite{qturbo2}:
\begin{itemize}
\item For a coded sequence of duration $N$, let $P_1 = [P_{1,1}, P_{1,2}, \dots, P_{1,t}, \dots, P_{1,N}]$ 
and $L_1 = [L_{1,1}, L_{1,2}, \dots, L_{1,t}, \dots, L_{1,N}]$, where $P_{1,t} \in G_n$ and $L_{1,t} \in G_k$. More explicitly, 
$P_{1,t} = [P_{1,t}^1, P_{1,t}^2, \dots, P_{1,t}^n ]$ and 
$L_{1,t} = [L_{1,t}^1, L_{1,t}^2, \dots, L_{1,t}^k ]$. For the ease of clarification, we will ignore the first subscript,
which represents the decoder, in the algorithm given below, i.e. we have $P_1 = P$ and $L_1 = L$. Similarly, $S_1^x = S^x$. 
 \item Let $U = (U_P:U_M)$, so that $U_P$ is the binary matrix formed by the first $2n$ columns of $U$ of \eqr{eq:cct_U}, while
$U_M$ is the binary matrix formed by the last $2m$ columns of $U$. Therefore, we have:
\begin{equation}
 P_t = \left(M_{t-1}:L_{t}:S_t\right)U_P.
\label{eq:Up}
\end{equation}
\begin{equation}
 M_t = \left(M_{t-1}:L_{t}:S_t\right)U_M.
\label{eq:Um}
\end{equation}
\item Let $\alpha_t\left(M_t\right)$ be the forward recursive coefficient, which is defined as follows:
\begin{align}
\alpha_t\left(M_t\right) &\triangleq \mathbf{P}\left(M_t|S^x_{\leq t}\right), \nonumber \\
& \propto \sum_{\mu, \lambda, \sigma} \mathbf{P}^a\left(L_{t} = \lambda\right) \mathbf{P}_{ch}\left(P_{t}\right) \alpha_{t-1}\left(\mu\right),
\label{eq:alpha}
\end{align}
where $\mu \in G_m$, $\lambda \in G_{k}$ and $\sigma \in G_{n-k}$, while $\sigma = \sigma_x + \sigma_z$, having $\sigma_x = S^x_t$. Furthermore, we have
$P_{t} = (\mu:\lambda:\sigma)U_P$ and
$M_t = \left(\mu:\lambda:\sigma \right)U_M$. 
The channel information $\mathbf{P}_{ch}\left(P_{t}\right)$ is computed assuming that each qubit
is independently transmitted over a quantum depolarizing channel having a depolarizing probability of $p$, whose channel transition probabilities
are given by~\cite{qturbo2}:
\begin{equation}
\mathbf{P}_{ch}\left(P_{t}^i\right) = \left\{
\begin{array}{l l}
 1-p, & \text{if } \mathcal{P}_t^i = \mathbf{I} \\
 p/3, & \text{if } \mathcal{P}_t^i \in \{\mathbf{X}, \mathbf{Z}, \mathbf{Y}\}. \\
\end{array}
\right.
\label{eq:dp_ch}
\end{equation} 
\item Let $\beta_{t}\left(M_{t}\right)$ be the backward recursive coefficient, which is defined as follows:
\begin{align}
& \beta_t\left(M_t\right) \triangleq \mathbf{P}\left(M_t|S^x_{> t}\right), \nonumber \\
& \propto \sum_{\lambda, \sigma} \mathbf{P}^a\left(L_{t} = \lambda\right) \mathbf{P}_{ch}\left(P_{t+1}\right) \beta_{t+1}\left(M_{t+1}\right),
\label{eq:beta}
\end{align}
where $P_{t+1} = (M_t:\lambda:\sigma)U_P$ and
$M_{t+1} = \left(M_t:\lambda:\sigma \right)U_M$.
\item Finally, we have the \textit{a-posteriori} probability $\mathbf{P}^o(L_{t})$, which is given by:
\begin{align}
 &\mathbf{P}^o(L_{t}) \triangleq \mathbf{P}(L_{t}|S^x), \nonumber \\
& \propto \sum_{\mu, \sigma} \mathbf{P}^a(L_t)\mathbf{P}_{ch}(P_t) \alpha_{t-1}\left(\mu\right) \beta_{t}\left(M_t\right),
\label{eq:app}
\end{align}
where $P_{t} = (\mu:L_{t}:\sigma)U_P$ and
$M_t = \left(\mu:L_{t}:\sigma \right)U_M$.
\item Marginalized probabilities $\mathbf{P}^o(L^{j}_{t})$ for $j \in \{0,k-1\}$ are then computed from $\mathbf{P}^o(L^{j}_{t})$ and the
\textit{a-priori} information is removed in order to yield the \textit{extrinsic} probabilities\cite{wilde_turbo2}, i.e we have:
\begin{equation}
 \ln [\mathbf{P}^e (L^{j}_{t})] = \ln [\mathbf{P}^o (L^{j}_{t})] - \ln [\mathbf{P}^a (L^{j}_{t})].
\label{eq:ext}
\end{equation}
\end{itemize}
Finally, the
\textit{extrinsic} average mutual information between $L_1$ and $\mathbf{P}^{e}_1(L_1)$ is computed, i.e. $I_E(L_1) = I[L_1,\mathbf{P}^{e}_1(L_1)]$.
Since the equivalent classical capacity of a quantum channel is given by the capacity achievable over each half of the $4$-ary symmetric channel as depicted
in \eqr{eq:c4},
$I_E(L_1)$ is the normalized mutual information of the $4$-ary symbols, which can be computed based on~\cite{kliewer:exit-symbol,mike3} as:
\begin{equation}
 I_E(L_1) = \frac{1}{2}\left(2 + \mbox{E} \left[ \sum_{\text{m} = 0}^{3} \mathbf{P}^{e}_1(L_1^{j(\text{m})}) \log_2\mathbf{P}^{e}_1(L_1^{j(\text{m})}) \right]\right),
\label{eq:MI}
\end{equation}
where \mbox{E} is the expectation (or time average) operator and $L_1^{j(\text{m})}$ is the $4$-ary $\text{m}^{\text{th}}$ hypothetical error imposed on the 
logical qubits. More explicitly, since error on each qubit is represented by an equivalent pair of classical bits, $L_1^{j(\text{m})}$ is a $4$-ary classical symbol 
with $\text{m} \in \{0,3\}$. The process is repeated for a range of $I_A(L_1) \in [0,1]$ values for obtaining the \textit{extrinsic} information transfer characteristics 
at the depolarizing probability $p$. The resultant inner EXIT function $T_{1}$ of the specific inner decoder may be defined as follows: 
\begin{equation}
 I_E(L_1) = T_{1}[I_A(L_1),p],
\label{eq:t1}
\end{equation}
which is dependent on the depolarizing probability $p$ of the quantum channel. 

\begin{figure}[tb]
  \begin{center}
  {\includegraphics[width=1\linewidth]{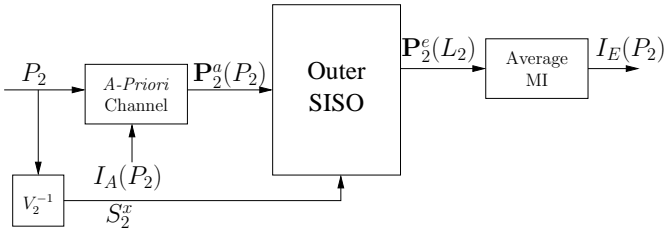}}
    \caption{System model for generating the EXIT chart of the outer decoder.}
  \label{fig:exit_outer}
  \end{center}
\end{figure}

The system model used for generating the EXIT chart of the outer decoder is depicted in \fref{fig:exit_outer}. As inferred from the figure, the EXIT curve
of the outer decoder is independent of the channel's output information. The \textit{a-priori} information is generated by the \textit{a-priori}
channel based on $P_2$ (error on the physical qubits of the second decoder) and $I_A(P_2)$, which is the average Mutual Information (MI)
between $P_2$ and $\mathbf{P}^a_2(P_2)$. Furthermore, as for the inner decoder, $P_2$ is passed through the inverse outer encoder $V_2^{-1}$ to
compute $S^x_2$, which is fed to the outer SISO decoder to yield the \textit{extrinsic} information $\mathbf{P}^e_2(P_2)$. Based on \eqr{eq:alpha} and \eqref{eq:beta},
this may be formulated as follows~\cite{qturbo2}:
\begin{align}
 &\mathbf{P}^o(P_{t}) \triangleq \mathbf{P}(P_{t}|S^x), \nonumber \\
& \propto \sum_{\mu, \lambda, \sigma} \mathbf{P}(P_t)\mathbf{P}(L_t=\lambda) \alpha_{t-1}\left(\mu\right) \beta_{t}\left(M_t\right),
\label{eq:app2}
\end{align}
where $P_{t} = (\mu:\lambda:\sigma)U_P$ and
$M_t = \left(\mu:\lambda:\sigma \right)U_M$. The resultant probabilities are marginalized and the \textit{a-priori} information is removed similar
to \eqr{eq:ext}.
The average MI between $P_2$ and 
$\mathbf{P}^e_2(P_2)$ is then calculated using \eqr{eq:MI}. The resultant EXIT chart is characterized by the following MI transfer function:
\begin{equation}
 I_E(P_2) = T_{2}[I_A(P_2)],
\label{eq:t2}
\end{equation}
where $T_{2}$ is the outer EXIT function, which is dependent on the specific outer decoder, but is independent of the depolarizing probability $p$.

Finally, the MI transfer characteristics of both decoders characterized by \eqr{eq:t1} and \eqr{eq:t2} are plotted in the same graph,
with the $x$ and $y$ axes of the outer decoder swapped. 
\section{Results and Discussions} \label{sec3}
\subsection{Accuracy of EXIT Chart Predictions} 
In order to verify the accuracy of our EXIT-chart based approach, we have analyzed the convergence behaviour of a rate-$1/9$ QTC, consisting of two identical
rate-$1/3$ QCCs. More specifically, for both the inner and outer decoders, we have used the configuration termed as ``PTO1R'' in~\cite{wilde_turbo,wilde_turbo2},
which is a non-catastrophic but quasi-recursive code. 

\begin{figure}[tb]
  \begin{center}
  {\includegraphics[scale = 0.7]{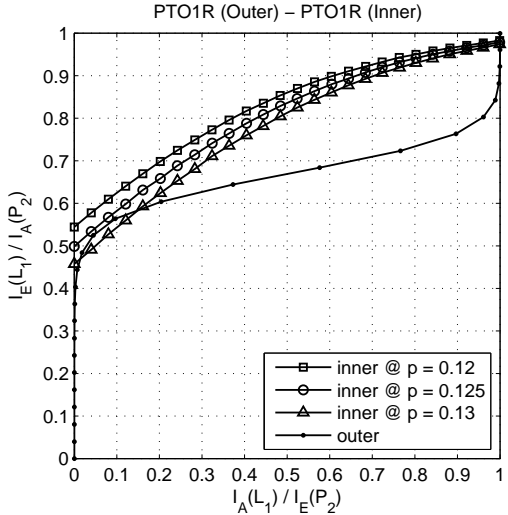}}
    \caption{The EXIT curves of a QTC parametrized by the increasing depolarizing probability $p$ \textit{(rate-1/9 QTC having PTO1R as both the inner
and outer components was used)}.}
  \label{fig:res_exit_inner}
  \end{center}
\end{figure}

Our first aim was to predict the convergence threshold using EXIT charts, which would otherwise require 
time-consuming Word Error Rate/Qubit Error Rate (WER/QBER) simulations. Convergence threshold 
can be determined by finding the maximum depolarizing probability $p$, which yields a marginally open EXIT tunnel between the EXIT curves
of the inner and outer decoder; hence, facilitating an infinitesimally low QBER. 
\fref{fig:res_exit_inner} shows the EXIT curves for the inner and outer decoders, where the area under the EXIT
curve of the inner decoder decreases upon increasing $p$. Eventually, the inner and outer curves crossover, when $p$ is increased to $p = 0.13$. More explicitly,
increasing $p$ beyond $0.125$, closes the EXIT tunnel. Hence, the convergence threshold is around $p = 0.125$.

\begin{figure}[tb]
  \begin{center}
  {\includegraphics[scale = 0.7]{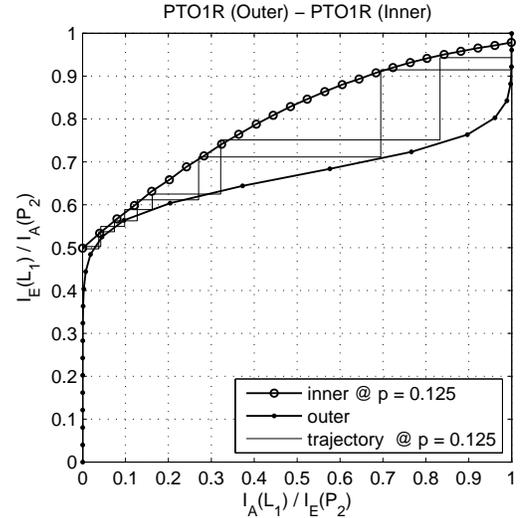}}
    \caption{The EXIT chart of a QTC with decoding trajectories at $p = 0.125$ \textit{(rate-1/9 QTC having PTO1R as both the inner
and outer components with an interleaver length of $30,000$ qubits was used)}.}
  \label{fig:exit_traj}
  \end{center}
\end{figure}

\fref{fig:exit_traj} shows two decoding trajectories superimposed on the EXIT chart of \fref{fig:res_exit_inner} at $p = 0.125$. We have used a $30,000$-qubit
long interleaver. As seen from \fref{fig:exit_traj}, the trajectory
successfully reaches the $(x,y) = (1,y)$ point of the EXIT chart. This in turn guarantees an infinitesimally low QBER at $p = 0.125$ for
an interleaver of infinite length. 

\begin{figure}[tb]
  \begin{center}
  {\includegraphics[scale = 0.65]{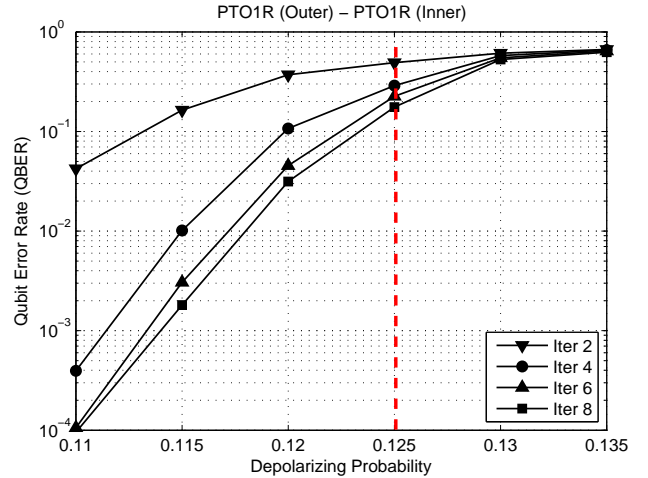}}
    \caption{QBER performance curve with increasing iteration number for an interleaver length of 3,000 qubits. \textit{Rate-1/9 QTC having PTO1R as both the 
inner and outer components was used.}}
  \label{fig:PTO1R_PTO1R_k_qber}
  \end{center}
\end{figure}

We have further verified the validity of our EXIT chart predictions using QBER simulations. \fref{fig:PTO1R_PTO1R_k_qber} shows
the QBER performance curve for an interleaver length of $3,000$ qubits. The performance improves upon increasing the number of iterations. More
specifically, the turbo-cliff region starts around $p = 0.125$, whereby the QBER drops as the iterations proceed. Therefore,
our EXIT chart predictions closely follow the Monte-Carlo simulation results. 
\subsection{Entanglement-Assisted and Unassisted Inner Codes} 
All non-catastrophic convolutional codes are non-recursive~\cite{qturbo2}. 
Therefore, the resultant families of QTCs have a bounded minimum distance and do not have a true iterative threshold.
To circumvent this limitation of QTCs, Wilde \textit{et al.}~\cite{wilde_turbo,wilde_turbo2} proposed to employ entanglement-assisted inner codes, which are
recursive as well as non-catastrophic. The resulting families of entanglement-assisted QTCs have an unbounded minimum distance~\cite{wilde_turbo,wilde_turbo2}, i.e. their 
minimum distance increases almost linearly with the interleaver length. Here, we verify this by analyzing the inner decoder's EXIT curves for both the unassisted (non-recursive) and entanglement-assisted
(recursive) inner convolutional codes.

\begin{figure}[tb]
  \begin{center}
  {\includegraphics[scale = 0.7]{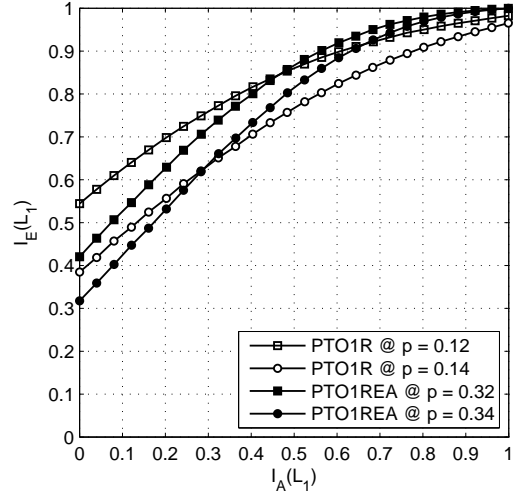}}
    \caption{ Comparison of the inner EXIT curves of both unassisted and entanglement-assisted QCCs, labeled as PTO1R and PTO1REA respectively.}
  \label{fig:PTO1R_PTO1REA_inner_comp}
  \end{center}
\end{figure}

For classical recursive inner codes, the inner decoder's EXIT curve reaches the $(x,y) = (1,1)$ point\footnote{Note that we only need $(x,y) = (1,y)$
for achieving decoding convergence to an infinitesimally low QBER. However, this requires an outer code having a sufficiently large minimum
distance for the sake of ensuring that the outer code's EXIT curve does not intersect with that of the inner code before reaching the $(1,y)$ point.
Unfortunately, an outer code having a large minimum distance would result in an
EXIT curve having in a large open-tunnel area. Thus, it will operate far from the capacity.}, which guarantees perfect decoding 
convergence to a vanishingly low QBER as well as having an unbounded minimum distance for the infinite family of QTCs~\cite{qturbo2} based on these inner codes. 
Consequently, the resulting families of QTCs have unbounded minimum distance
and hence an arbitrarily low QBER can be achieved for an infinitely long interleaver. This also holds true for recursive quantum convolutional codes, as shown in
\fref{fig:PTO1R_PTO1REA_inner_comp}. In this figure, we compare the inner decoder's EXIT curves of both the unassisted and the entanglement-assisted
QCCs of~\cite{wilde_turbo}, which are labeled ``PTO1R'' and ``PTO1REA'', respectively. For the PTO1R configuration, decreasing the depolarizing probability from $p = 0.14$
to $p = 0.12$ shifts the inner decoder's EXIT curve upwards and towards the $(1,1)$ point. Hence, the EXIT curve will 
manage to reach the $(1,1)$ point only at very low values of depolarizing probability. By contrast, the EXIT curve of PTO1REA always terminates at $(1,1)$, regardless 
of the value of $p$. Therefore, provided an open EXIT
tunnel exists and the interleaver length is sufficiently long, the decoding trajectories of an entanglement-assisted QTC will always reach the $(1,1)$ point; thus,
guaranteeing an arbitrarily low QBER for the infinite family of QTCs based on these inner codes. 
In other words, the performance improves upon increasing the interleaver length; thus, implying that the minimum distance
increases upon increasing the interleaver length and therefore the resultant QTCs have an unbounded minimum distance.
\subsection{Optimized Quantum Turbo Code Design} 
The QTC design of~\cite{wilde_turbo,wilde_turbo2} characterized in \fref{fig:exit_traj} exhibits a large area between the inner and outer decoder's EXIT curves.
The larger the `open-tunnel' area, the farther the QBER performance curve from the achievable capacity limit~\cite{tc_teq_st_2:book}. Consequently, various distance spectra based QTCs 
investigated in~\cite{wilde_turbo2} operate within $0.9$ dB of the hashing bound. For the sake of achieving a near-capacity performance, we minimize the 
area between the inner and outer EXIT curves, so that a narrow, but still marginally open tunnel exists at the highest possible depolarizing probability.
Our aim was to construct a rate-$1/9$ QTC relying on an entanglement-assisted inner code (recursive and non-catastrophic) and an unassisted outer code
(non-catastrophic) having a memory of $3$ and a rate of $1/3$. 
The resultant QTC has an entanglement consumption rate of $6/9$, for which the corresponding maximum tolerable depolarizing probability
was shown to be $p_{\max} = 0.3779$ in~\cite{wilde_turbo2}. 

\begin{figure}[tb]
  \begin{center}
  {\includegraphics[scale = 0.7]{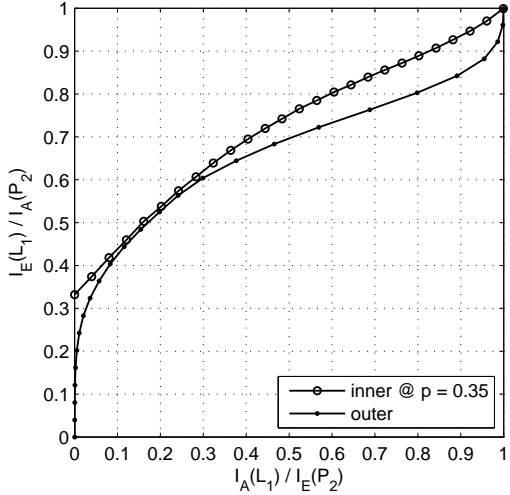}}
    \caption{EXIT chart of the optimized rate-$1/9$ QTC \textit{(Interleaver length = $30,000$ qubits)}.}
  \label{fig:exit_traj_opt_1}
  \end{center}
\end{figure}

\begin{figure}[tb]
  \begin{center}
  {\includegraphics[scale = 0.65]{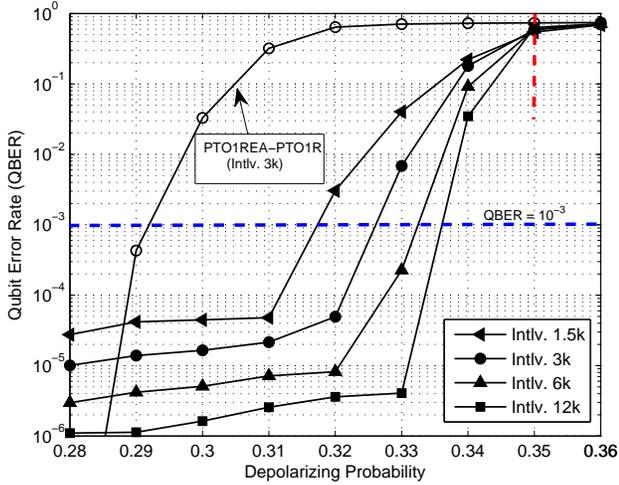}}
    \caption{QBER performance curves of the optimized rate-$1/9$ QTC for varying interleaver lengths and a maximum of $15$ iterations.}
  \label{fig:opt_wer_ber_1}
  \end{center}
\end{figure}

For the sake of designing a near-capacity QTC operating close to the capacity limit of $p_{\max} = 0.3779$, we randomly selected both inner and outer encoders 
from the Clifford group according to the algorithm 
of~\cite{Divincenzo2002} in order to find the inner and outer components, which minimize the area between the corresponding EXIT curves. 
Based on this design criterion, we found optimal inner and outer code pair whose seed transforms\footnote{Please refer to~\cite{wilde_turbo2} for the details of this 
representation.} (decimal representation) are given by:
\begin{align}
 U_{\text{inner}} = \{&4091, 3736, 2097, 1336, 1601, 279, \nonumber \\
                  &3093, 502, 1792, 3020, 226, 1100\};
\label{eq:uinner}
\end{align} 
\begin{align}
 U_{\text{outer}} = \{&1048, 3872, 3485, 2054, 983, 3164, \nonumber \\
                  &3145, 1824, 987, 3282, 2505, 1984\}.
\end{align}
\fref{fig:exit_traj_opt_1} shows the corresponding EXIT chart at the convergence threshold of $p = 0.35$. As observed
in \fref{fig:exit_traj_opt_1}, a marginally open EXIT tunnel exists
between the two curves, which facilitates for the decoding trajectories to reach the $(1,1)$ point. 
Hence, our optimized QTC has a convergence threshold of $p = 0.35$, which is only $\left[10 \times \log_{10} (\frac{0.35}{0.3779})\right] = 0.3$ dB from the maximum tolerable depolarizing probability of $0.3779$.
The corresponding QBER performance curves recorded for our optimized design are 
given in \fref{fig:opt_wer_ber_1}. A maximum of $15$ iterations were used, while the interleaver length was increased from $1500$ to $12,000$. Similar to classical
turbo codes, increasing the interleaver length for $p < 0.35$ improves the attainable performance.
Furthermore, \fref{fig:opt_wer_ber_1} also compares our optimized design with the rate-$1/9$ QTC of~\cite{wilde_turbo2} for an interleaver length of $3000$,
which is labeled ``PTO1REA-PTO1R'' in the figure. For the ``PTO1REA-PTO1R'' configuration, the turbo cliff region emerges around $0.31$, which is within
$0.9$ dB of the capacity limit.
Therefore, our EXIT-chart based QTC outperforms the QTC design based on the distance spectrum~\cite{wilde_turbo2}. 
More specifically, the ``PTO1REA-PTO1R'' configuration yields a QBER of $10^{-3}$ at $p = 0.2925$, while our optimized QTC
gives a QBER of $10^{-3}$ at $p = 0.3275$. Hence, our optimized QTC outperforms the `PTO1REA-PTO1R'' configuration by about 
$\left[10 \times \log_{10} (\frac{0.2925}{0.3275})\right] = 0.5$ dB at a QBER of $10^{-3}$.
However, our main design objective was to not to carry out an exhaustive code search, but to demonstrate the explicit benefit of our EXIT-chart based approach
in the context of quantum codes.
It must also be observed in \fref{fig:opt_wer_ber_1} that a relatively high error floor
exists for our optimized design, which is gradually reduced upon increasing the interleaver length. This is because the outer code
has a low minimum distance of only $3$. Its truncated distance spectrum is as follows:
\begin{align}
 D(x) = &2x^3 + 19x^4 + 108x^5 + 530x^6 + 2882x^7 + 14179x^8 + \nonumber \\
 &62288x^9 + 243234x^{10} + 845863x^{11} + 1165784x^{12} + \nonumber \\
 &2501507x^{13} + 744394x^{14}. \nonumber
\end{align}
By contrast, the truncated distance spectrum of ``PTO1R'', which has a minimum distance of $5$, is given by~\cite{wilde_turbo2}:
\begin{align}
 D(x) = &11x^5 + 47x^6 + 253x^7 + 1187x^8 + 6024x^9 + \nonumber \\
 &30529x^{10} + 153051x^{11} + 771650x^{12}.\nonumber
 \end{align}
Consequently, as gleaned from \fref{fig:opt_wer_ber_1}, the ``PTO1REA-PTO1R'' configuration has a much lower error floor ($< 10^{-6}$), 
since the outer code ``PTO1R'' has a higher minimum distance.
However, this enlarges the area between the inner and outer decoder's EXIT curves; thus, driving the performance farther away from the achievable capacity,
as depicted in~\fref{fig:exit_traj}.
Hence, there is a trade-off between the minimization of the error floor and achieving a near-capacity performance. More specifically,
while the distance-spectrum based design primarily aims for achieving a lower error floor, the EXIT-chart based design strives for 
achieving a near-capacity performance. 
\section{Conclusions} \label{sec4} 
In this contribution, we have extended the application of classical non-binary EXIT charts 
to the circuit-based syndrome decoder of quantum turbo codes, in order 
to facilitate the EXIT-chart based design of QTCs. We have verified the accuracy of our EXIT chart generation approach
by comparing the convergence threshold predicted by the EXIT chart to the Monte-Carlo simulation results. 
Furthermore, we have shown with the aid of EXIT charts that entanglement-assisted recursive QCCs have
an unbounded minimum distance.
Moreover, we have designed an optimal entanglement-assisted QTC using EXIT charts, which outperforms the distance spectra based QTC of~\cite{wilde_turbo2}
by about $0.5$ dB at a QBER of $10^{-3}$.
\section*{Acknowledgement}
We would like to thank Dr. Mark M. Wilde for the valuable discussions. 


%
\bibliographystyle{IEEEtran}
\bibliography{QTC_Exit-2col}

\end{document}